# Synthesizing beta-amyloid PET images from T1-weighted Structural MRI: A Preliminary Study

Qing Lyu, Jin Young Kim, Jeongchul Kim, and Christopher T Whitlow

Wake Forest University School of Medicine, Winston-Salem NC 27101, USA
`qlyu@wakehealth.edu`

**Abstract.** Beta-amyloid positron emission tomography (Aβ-PET) imaging has become a critical tool in Alzheimer's disease (AD) research and diagnosis, providing insights into the pathological accumulation of amyloid plaques, one of the hallmarks of AD. However, the high cost, limited availability, and exposure to radioactivity restrict the widespread use of Aβ-PET imaging, leading to a scarcity of comprehensive datasets. Previous studies have suggested that structural magnetic resonance imaging (MRI), which is more readily available, may serve as a viable alternative for synthesizing Aβ-PET images. In this study, we propose an approach to utilize 3D diffusion models to synthesize Aβ-PET images from T1-weighted MRI scans, aiming to overcome the limitations associated with direct PET imaging. Our method generates high-quality Aβ-PET images for cognitive normal cases, although it is less effective for mild cognitive impairment (MCI) patients due to the variability in Aβ deposition patterns among subjects. Our preliminary results suggest that incorporating additional data, such as a larger sample of MCI cases and multi-modality information including clinical and demographic details, cognitive and functional assessments, and longitudinal data, may be necessary to improve Aβ-PET image synthesis for MCI patients.

**Keywords:** Magnetic Resonance Imaging, Positron Emission Tomography, Deep Learning, Medical Image Synthesis, Diffusion Model.

## 1    Introduction

Alzheimer's disease (AD) is a progressive and degenerative brain disorder and the most common form of dementia in older adults. Currently, AD is recognized as the sixth-leading cause of death in the United States [1]. In 2024, an estimated 6.9 million Americans aged 65 and older are living with Alzheimer's dementia [1]. AD is marked by changes in the brain, notably the excessive accumulation of beta-amyloid (Aβ) protein fragments and abnormal forms of the protein tau [1]. Recent advancements in radiotracers that bind to Aβ plaques and paired helical filaments of tau, which forms neurofibrillary tangles, have enabled the visualization and quantification of AD pathology in living patients through positron emission tomography (PET) scans [2, 3]. However, a major challenge in AD research is the scarcity of PET scan data. Compared to magnetic resonance imaging (MRI), a more commonly used imaging technique, PET scans pose a higher health risk due to ionizing radiation, which can increase the likelihood of cancer. Additionally, PET scans are more expensive and less widely available, limiting their use in many medical centers.

To address the challenges posed by the limited availability of PET data, researchers are exploring the use of artificial intelligence to synthesize PET images from other medical imaging modalities, such as MRI. This data-driven approach aims to learn the complex, non-linear mappings between these different domains. Various deep learning algorithms have been proposed, including convolutional neural networks (CNN), generative adversarial networks (GAN), transformers, and diffusion models. Sikka et al. introduced a 3D UNet model to generate Fludeoxyglucose F18 (FDG) PET images from MRI data [4]. Building on this, Emami et al. developed a frequency-aware attention UNet model to enhance FDG-PET synthesis from MRI [5]. In the realm of GANs, Sikka et al. also proposed a global and local-aware GAN for creating FDG-PET images [6], while Yan et al. employed a conditional GAN to synthesize Aβ-PET images [7]. Wei et al. introduced a sketcher-refiner generative model for Aβ-PET synthesis from MRI [8], and both Hu et al. and Lin et al. developed bidirectional mapping GANs to generate FDG-PET images from MRI [9, 10]. Additionally, Shin et al. implemented a bidirectional encoder representations from transformers (BERT) algorithm to generate Aβ-PET from MRI [11]. Vision transformers have also been investigated for PET synthesis. For instance, Zhang et al. designed a spatial adaptive and transformer fusion network to synthesize full dose FDG-PET images from MRI and low dose FDG PET scans [12]. Following the success of diffusion models in natural image generation, researchers have started applying them to PET synthesis. Xie et al. proposed a joint diffusion model to convert high-field MRI into FDG-PET images [13], while Jang et al. developed a text-guided image synthesis technique capable of generating realistic tau-PET images from textual descriptions and MRI data [14].

Although various studies have explored the synthesis of PET images from MRI, the majority have focused on FDG-PET, which measures glucose metabolism in the brain and reflects the activity of brain cells. In contrast, the synthesis of Aβ-PET and tau-PET, both of which are more directly related to AD as they target the specific protein abnormalities associated with the condition, are less studied. Moreover, there is limited analysis of the disparities in synthetic PET results across different groups of AD patients, such as those at various stages of the disease. This gap in research makes it challenging to accurately assess the quality of synthetic PET images and to use them effectively in AD-related downstream studies.

In this study, we propose a 3D diffusion model to synthesize Aβ-PET images from corresponding MRI scans and evaluate its performance across different AD groups. The main contributions of our study are as follows:

1. We employed a 3D diffusion model for Aβ-PET synthesis, leveraging both intra- and inter-plane information from MRI images. This 3D approach mitigates the issue of pixel intensity mismatches across adjacent slices, a common problem in 2D models.

2. We focused specifically on synthesizing Aβ-PET images, which are directly linked to AD pathology but have been less explored in previous studies.

3. We evaluated the quality of the synthetic images across various AD stages. Our model performed well in cognitive normal patients, likely due to the relatively consistent Aβ distribution pattern within this group. However, it performed less effectively in early- and late-stage mild cognitive impairment (MCI) patients, where Aβ distribution shows greater heterogeneity. Our findings suggest that additional information, beyond MRI scans, may be necessary to improve the synthetic Aβ-PET images in more advanced AD stages.

## 2 Methodology

### 2.1 Dataset

We utilized data from the Alzheimer's Disease Neuroimaging Initiative (ADNI) database, comprising 108 cognitive normal (CN) subjects, 163 early mild cognitive impairment (EMCI) patients, and 80 late mild cognitive impairment (LMCI) patients. MRI and Aβ-PET scans were paired based on their acquisition dates, and the Advanced Normalization Tools (https://github.com/ANTsX/ANTs) were used to register the data to the MNI152 template. After preprocessing, a total of 838 MRI and Aβ-PET scan pairs were involved in this study. Of these, 168 pairs were randomly selected for validation and testing, while the remaining 670 pairs were used for model training.

### 2.2 Dataset

The denoising diffusion probabilistic model (DDPM) was employed for PET synthesis [15]. DDPM consists of two main processes: a forward diffusion process and a backward denoising process. The forward diffusion process can be formulated as follows:

$$q(x_t|x_{t-1}) = \mathcal{N}(x_t; \sqrt{1-\beta_t} \cdot x_{t-1}, \beta_t \cdot I), \qquad (1)$$

$$q(x_{1:T}|x_0) = \prod_{t=1}^{T} q(x_t|x_{t-1}), \qquad (2)$$

where $T$ is the total number of noising steps, $\beta_t \in (0,1)$ controls the variance of incremental Gaussian noise, and $\mathcal{N}(x; \mu, \sigma)$ represents a Gaussian distribution of mean $\mu$ and variance $\sigma$. For the backward denoising process, a neural network is trained to approximate $q(x_{t-1}|x_t)$ in each step, and estimate the mean $\mu_\theta(x_t, t)$ and the variance $\Sigma_\theta(x_t, t)$:

$$q(x_{t-1}|x_t, x_0) = \mathcal{N}(x_{t-1}; \mu_\theta(x_t,t), \tilde{\beta}_t \cdot I), \qquad (3)$$

$$\mu_\theta(x_t,t) = \frac{1}{\sqrt{\alpha_t}}(x_t - \frac{1-\alpha_t}{\sqrt{1-\bar{\alpha}_t}}\epsilon_\theta(x_t,t)), \qquad (4)$$

$$\tilde{\beta}_t = \frac{1-\bar{\alpha}_{t-1}}{1-\bar{\alpha}_t}\beta_t, \qquad (5)$$

where $\alpha_t = 1 - \beta_t$, and $\bar{\alpha}_t = \prod_{i=1}^{t} \alpha_i$.

In this study, we incorporated MRI images as a condition to guide the content of the generation results. The objective function can be express as follows:

$$\mathcal{L} = \mathbb{E}_{t \sim [1,T], x_0, \epsilon_t}[\| \epsilon_t - \epsilon_\theta(\sqrt{\bar{\alpha}_t}x_0 + \sqrt{1-\bar{\alpha}_t}\,\epsilon_t, y, t)\|^2], \qquad (6)$$

where $x_0$ is the target PET images, $y$ is the MRI conditional images, $\epsilon_t \in \mathcal{N}(0, \mathrm{I})$.

UNet was employed as the denoising model in the DDPM, consisting of four down-sampling and four up-sampling blocks. After each down-sampling or up-sampling block, feature maps are either down-sampled or up-sampled by a factor of two. Each block contains two residual units, with each unit comprising two 3D convolutional layers, followed by group normalization and a Swish activation function.

### 2.3 Implementation details

The number of channels in each UNet down-sampling block are 128, 256, 512, and 512, respectively. A 3D convolution was applied using a kernel size of 3×3×3, with a stride of 1 and padding of 1. During training, the input data were randomly cropped into multiple 64×64×64 patches and fed into the neural network. The training process was carried out over 2,500 epochs. For inference, the entire image volume was input into the neural network. All experiments were conducted on an Nvidia H100 GPU card.

## 3 Result

### 3.1 Aβ-PET synthesis results

We present synthetic results from a CN case, an EMCI case, and a LMCI case in Fig. 1. The synthetic PET results closely resemble the real PET images for the CN case, with only minor errors, whereas greater discrepancies are observed in the EMCI and LMCI cases. Table 1 shows the statistical evaluation in terms of structural similarity index measure (SSIM) and peak signal-to-noise ratio (PSNR). The CN results exhibit significantly higher performance compared to the EMCI and LMCI cases.

**Table 1.** 95% confidence interval of Aβ-PET synthesis results.

|       | SSIM              | PSNR                |
| ----- | ----------------- | ------------------- |
| CN    | $0.911 \pm 0.014$ | $24.976 \pm 0.856$  |
| EMCI  | $0.855 \pm 0.013$ | $21.505 \pm 0.814$  |
| LMCI  | $0.840 \pm 0.014$ | $19.734 \pm 0.651$  |

### 3.2 Aβ-deposition heterogeneity comparison across different AD groups

We calculated the standard deviation of each voxel's standardized uptake value ratio (SUVR) value across CN, EMCI, and LMCI groups, based on all data included in this study. As shown in Fig. 2, the standard deviation of SUVR values is slightly higher for the LMCI group, particularly in the occipital region, as indicated by the red arrows.

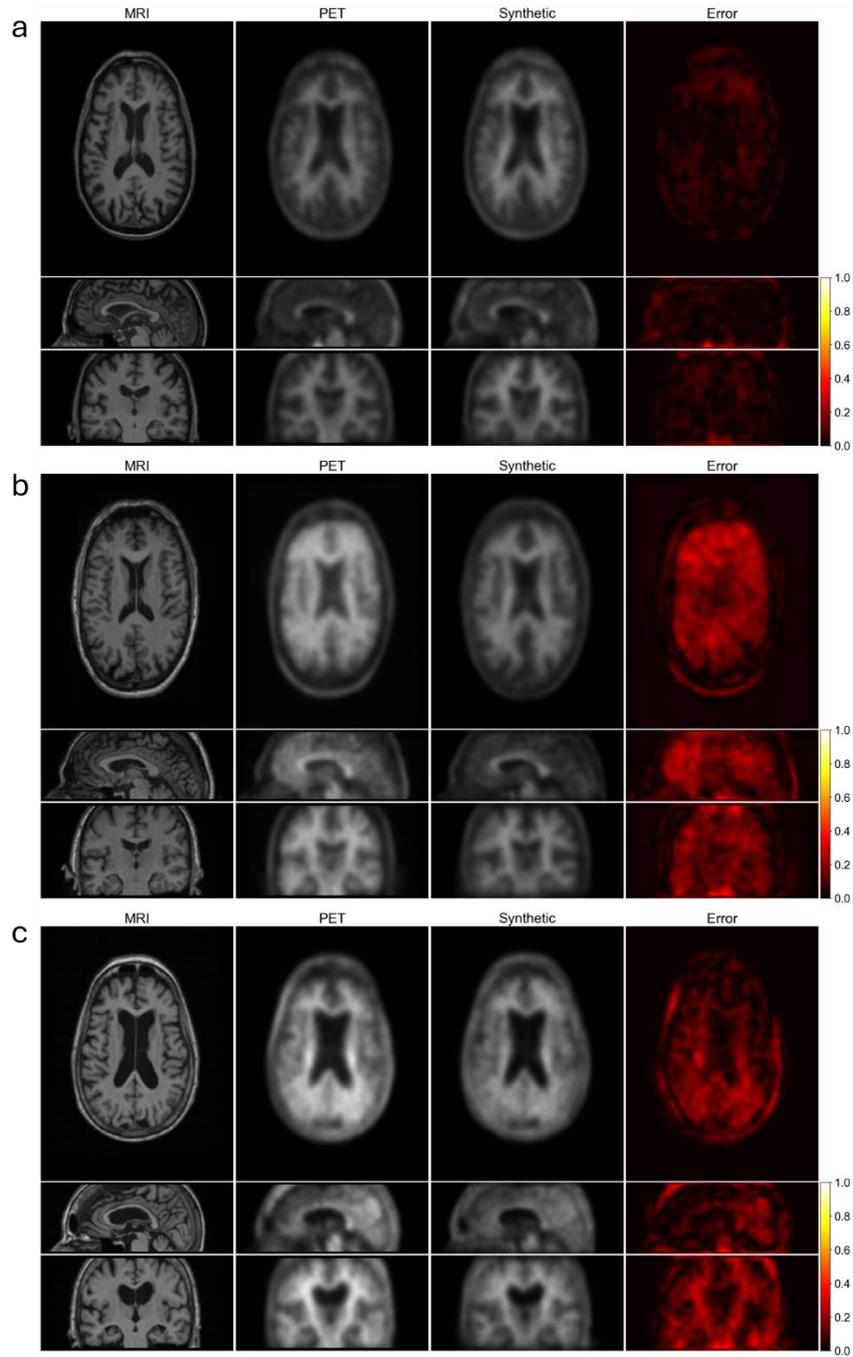

**Fig. 1.** Aβ-PET synthesis results. a) Cognitive normal case. b) Early-MCI case. c) Late-MCI case.

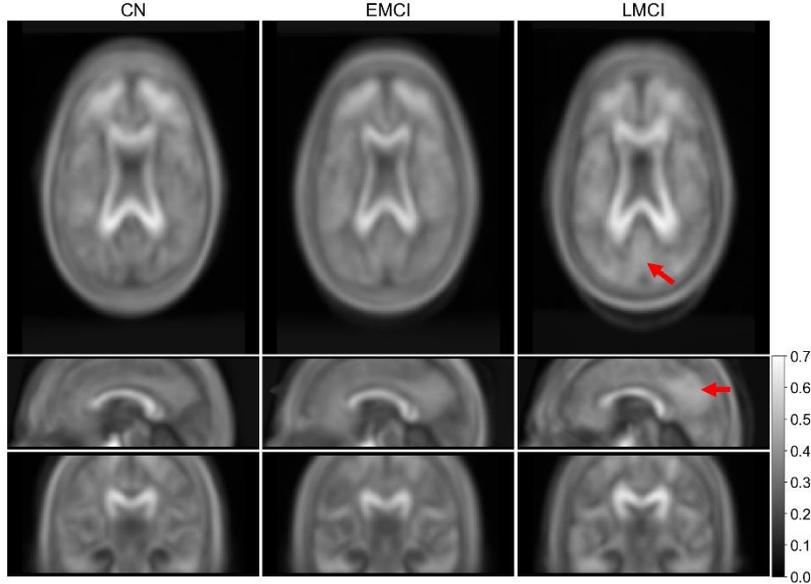

**Fig. 2.** Comparison of voxel-based SUVR standard deviations across CN, EMCI, and LMCI groups.

## 4 Discussion

Aβ-PET pathology is among the earliest detectable brain changes in AD pathogenesis [16]. As a result, Aβ-PET imaging has become an essential tool for AD research with the advantage of detecting and quantifying amyloid plaques in the living brain. However, several limitations, including high scan costs, limited availability of scanners, and exposure to harmful radiation, result in limited Aβ-PET imaging data, restricting its broader application in AD research. Recent studies suggest that data-drive generative AI algorithms can effectively learn the complex, nonlinear relationships between structural MRI and Aβ-PET, enabling the synthesis of Aβ-PET images from MRI [17]. Despite this potential, most existing research has focused on FDG-PET synthesis, leaving a gap in Aβ-PET synthesis studies. To address this, we propose a 3D diffusion model approach for Aβ-PET synthesis. Our method demonstrates strong performance in generating CN images and effectively alleviates pixel-intensity inconsistencies across adjacent slices, a common issue encountered in 2D generative models. Our results on CN synthesis confirm findings from previous studies, demonstrating that high-quality Aβ-PET images can be effectively generated from structural MRI.

We analyzed the standard deviation of pixel intensity in SUVR Aβ-PET images across CN, EMCI, and LMCI groups. Our results showed that Aβ accumulation in the brain exhibits greater heterogeneity in MCI patients compared to CN subjects. This finding may explain why our model did not perform as well in MCI patients. Compared to CN subjects, The increased heterogeneity in Aβ intracranial deposition among MCI patients suggests that they present more variable Aβ distribution patterns. Consequently, this variability makes it more challenging for AI models to learn the mapping between MRI and PET images effectively.

Our results suggest that additional information may be necessary to obtain high-quality Aβ-PET images for MCI patients. One potential solution is to expand the training dataset to include more MCI cases. For this purpose, resources such as the National Alzheimer's Coordinating Center (NACC) database can be utilized [18]. Another approach is to incorporate multi-modality data. Clinical and demographic information, including age, sex, and APOE genotype, cognitive and functional assessments, and longitudinal data could provide valuable supplementary information to improve synthesis outcomes.

# References

1. *2024 Alzheimer's disease facts and figures.* Alzheimers Dement, 2024. **20**(5): p. 3708-3821.
2. Klunk, W.E., et al., *Imaging brain amyloid in Alzheimer's disease with Pittsburgh Compound-B.* Ann Neurol, 2004. **55**(3): p. 306-19.
3. Chien, D.T., et al., *Early clinical PET imaging results with the novel PHF-tau radioligand [F-18]-T807.* J Alzheimers Dis, 2013. **34**(2): p. 457-68.
4. Sikka, A., S.V. Peri, and D.R. Bathula. *MRI to FDG-PET: Cross-Modal Synthesis Using 3D U-Net for Multi-modal Alzheimer's Classification.* 2018. Cham: Springer International Publishing.
5. Emami, H., Q. Liu, and M. Dong *FREA-Unet: Frequency-aware U-net for Modality Transfer.* 2020. arXiv:2012.15397 DOI: 10.48550/arXiv.2012.15397.
6. Sikka, A., et al. *MRI to PET Cross-Modality Translation using Globally and Locally Aware GAN (GLA-GAN) for Multi-Modal Diagnosis of Alzheimer's Disease.* 2021. arXiv:2108.02160 DOI: 10.48550/arXiv.2108.02160.
7. Yan, Y., et al. *Generation of Amyloid PET Images via Conditional Adversarial Training for Predicting Progression to Alzheimer's Disease.* 2018. Cham: Springer International Publishing.
8. Wei, W., et al., *Predicting PET-derived demyelination from multimodal MRI using sketcher-refiner adversarial training for multiple sclerosis.* Med Image Anal, 2019. **58**: p. 101546.
9. Lin, W., et al., *Bidirectional Mapping of Brain MRI and PET With 3D Reversible GAN for the Diagnosis of Alzheimer's Disease.* Front Neurosci, 2021. **15**: p. 646013.
10. Hu, S., et al., *Bidirectional Mapping Generative Adversarial Networks for Brain MR to PET Synthesis.* IEEE Trans Med Imaging, 2022. **41**(1): p. 145-157.
11. Shin, H.-C., et al. *GANBERT: Generative Adversarial Networks with Bidirectional Encoder Representations from Transformers for MRI to PET synthesis.* 2020. arXiv:2008.04393 DOI: 10.48550/arXiv.2008.04393.
12. Zhang, L., et al., *Spatial adaptive and transformer fusion network (STFNet) for low-count PET blind denoising with MRI.* Med Phys, 2022. **49**(1): p. 343-356.
13. Xie, T., et al. *Synthesizing PET images from High-field and Ultra-high-field MR images Using Joint Diffusion Attention Model.* 2023. arXiv:2305.03901 DOI: 10.48550/arXiv.2305.03901.

14. Jang, S.-I., et al. *TauPETGen: Text-Conditional Tau PET Image Synthesis Based on Latent Diffusion Models*. 2023. arXiv:2306.11984 DOI: 10.48550/arXiv.2306.11984.

15. Ho, J., A. Jain, and P. Abbeel, *Denoising diffusion probabilistic models*, in *Proceedings of the 34th International Conference on Neural Information Processing Systems*. 2020, Curran Associates Inc.: Vancouver, BC, Canada. p. Article 574.

16. Jack, C.R., Jr., et al., *Tracking pathophysiological processes in Alzheimer's disease: an updated hypothetical model of dynamic biomarkers.* Lancet Neurol, 2013. **12**(2): p. 207-16.

17. Dayarathna, S., et al., *Deep learning based synthesis of MRI, CT and PET: Review and analysis.* Med Image Anal, 2024. **92**: p. 103046.

18. Beekly, D.L., et al., *The National Alzheimer's Coordinating Center (NACC) database: the Uniform Data Set.* Alzheimer Dis Assoc Disord, 2007. **21**(3): p. 249-58.